\documentclass[12pt]{iopart}

\usepackage{iopams}  
\usepackage{graphicx}
\begin{document}

\title[]{ Perpendicular momentum injection by lower hybrid wave in a tokamak}

\author{Jungpyo Lee}
\address{Plasma Science and Fusion Center, MIT, Cambridge, USA}
\ead{Jungpyo@mit.edu}

\author{Felix I. Parra}
\address{Plasma Science and Fusion Center, MIT, Cambridge, USA}
\author{Ron R. Parker}
\address{Plasma Science and Fusion Center, MIT, Cambridge, USA}

\author{Paul T. Bonoli}
\address{Plasma Science and Fusion Center, MIT, Cambridge, USA}

\begin{abstract}
The injection of lower hybrid waves for current drive into a tokamak affects the profile of intrinsic rotation. In this article, the momentum deposition by the lower hybrid wave on the electrons is studied. Due to the increase in the poloidal momentum of the wave as it propagates into the tokamak, the parallel momentum of the wave increases considerably. The change of the perpendicular momentum of the wave is such that the toroidal angular momentum of the wave is conserved. If the perpendicular momentum transfer via electron Landau damping is ignored, the transfer of the toroidal angular momentum to the plasma will be larger than the injected toroidal angular momentum. A proper quasilinear treatment proves that both perpendicular and parallel momentum are transferred to the electrons. The toroidal angular momentum of the electrons is then transferred to the ions via different mechanisms for the parallel and perpendicular momentum. The perpendicular momentum is transferred to ions through an outward radial electron pinch, while the parallel momentum is transferred through collisions.
\end{abstract}

\pacs{52.30.-q,52.35.Hr,52.55}
\maketitle

\section{Introduction}
The momentum of radio-frequency (RF) waves has been studied since the early days of development of electromagnetics \cite{Bers:1972}. Recently, the experimental observation of plasma flows generated by RF waves has renewed the interest in momentum deposition by RF waves. For example, a significant ion toroidal rotation ($\sim$50km/s) has been measured by X-Ray spectroscopy for impurities in Alcator C-Mod during lower hybrid (LH) wave power injection \cite{Cushman:2009}, and the relation between the computed toroidal angular momentum input from LH waves and the measured initial change of ion toroidal rotation has been investigated \cite{Lee:RFC2011}. \\
\\
In tokamaks, the lower hybrid wave is used to drive plasma parallel current with asymmetric antenna spectra along the direction parallel to the static magnetic field {\cite{Fisch:RMP1987}. Due to electron Landau damping, the wave power is transferred to non-thermal fast electrons ($v_{\|}\sim 3v_{te}-10v_{te}$, where $v_{\|}$ is the parallel electron velocity and $v_{te}$ is the electron thermal velocity) {\cite{Fisch:RMP1987,Kennel:POF1966,Landau:JPUSSR1946}}. The toroidal phase velocity of the wave is chosen to increase current drive efficiency and ensure the accessibility of the wave to the core of the tokamak. The electron Landau damping can be described kinetically as a quasilinear velocity diffusion coefficient if the strength and the spectrum of the electric field satisfy some conditions given in \cite{Chirikov:PR1979,Laval:PPCF1999,LEE:POP2011}. Many of the observations (e.g. a driven current density and a hard X-ray diagnostic) in LH current drive experiments are well-reproduced by theory and simulation  \cite{Wright:POP2009,Meneghini:POP2009}. 
\\ \\
Recently, for high plasma densities (e.g. line averaged density $> 10^{20} \textrm{m}^{-3}$ in Alcator C-Mod), it has been observed that penetration of LH waves into the plasma core becomes problematic in many experiments in diverted tokamaks (e.g. Alcator C-Mod \cite{Wallace:POP2010}, FTU \cite{Cesario:Nature2010} and JET \cite{Cesario:PPCF2011}). The observation has motivated research on parasitic absorption mechanisms of the LH wave in the scrape-off-layer (SOL), such as collisional absorption \cite{Wallace:POP2010} and parametric decay instability  \cite{Cesario:Nature2010, Cesario:PPCF2011}. Modifying the edge electron temperature based on the theoretical prediction was found to be useful to overcome the density limit in FTU \cite{Cesario:Nature2010}. The observation of the significant ion toroidal rotation change due to the lower hybrid wave has been reported only in the low or medium density regime  (e.g. line averaged density $<10^{20} \textrm{m}^{-3}$ in Alcator C-Mod) \cite{Cushman:2009}, and for this reason, the coupling problems for high densities are not treated in this paper. This work is concerned only with LH wave momentum transfer in the core when the wave couples well and penetrates into the core.\\ \\
In this paper, we investigate how the wave momentum changes as the wave propagates from the launcher to the core of the tokamak where it is damped and transferred to the plasma. Especially, we focus on the toroidal angular momentum transfer to the electrons that is essential to explain the temporal behavior of the ion toroidal rotation initiated by the LH wave injection. We also investigate how the transferred momentum affects the radial motion of the electrons and ions. The radial particle pinch can be a channel to transfer the toroidal momentum to the ions. On the transport time scale, the ion turbulent momentum transport dominates the temporal evolution of the ion toroidal rotation \cite{Lee:RFC2011, Parra:PPCF2010}, but it is beyond the scope of this paper. Before studying this long time scale behavior we need to understand the momentum deposition.\\ \\ 
The wave momentum density is defined as $\mathbf{k}/\omega$ times the energy density, where $\mathbf{k}$ is the wave vector and $\omega$ is the wave frequency \cite{Bers:1972}. When the wave has a non-resonant interaction with the particles in a long propagation distance, the wave energy density does not change but the poloidal wave vector changes due to the dispersion relation of the lower hybrid wave, and consequently the poloidal wave momentum varies. On the other hand, when the wave has a resonant interaction with the particles in a short propagation distance, the wave energy density is reduced and the wave vector remains unchanged.\\ \\
Figure 1 shows the typical behavior of a LH wave in an inhomogeneous tokamak. As the wave propagates from the low field side launcher, it develops a very high poloidal wave vector (about 10 times larger than toroidal wave vector) due to the plasma dispersion relation \cite{Bonoli:PF1986}. The large poloidal wave vector contributes to the parallel wave number $k_\|$ as much as the toroidal wave vector does, even overcoming the small ratio of the poloidal magnetic field over the toroidal magnetic field, $B_\theta/B_\phi \sim 0.1$ ($k_\| = \frac{B_{\phi}}{B}k_\phi+\frac{B_\theta}{B}k_\theta$ in a circular tokamak, where $B_\phi$, $B_\theta$, $B$ are toroidal, poloidal, and total magnetic field, and $k_\phi$, $k_\theta$ are toroidal and poloidal wave numbers, respectively). That results in the parallel refractive index $n_\| \equiv k_\| c/\omega \sim -3$ of the damped wave, significantly larger than the toroidal index $n_\phi \equiv k_\phi c/\omega \sim -1.6$ at the launcher, as shown in Fig. 1 (the negative sign means that the wave propagates in the counter-current direction of the tokamak). Here, $c$ is the speed of light. The electron Landau damping of the wave becomes stronger where the phase velocity of the wave becomes lower (in other words, where the refractive index becomes higher), since a lower phase velocity resonates with more electrons.\\ \\ 
As shown in Fig. 1, until the wave reaches the region where the parallel phase velocity of the wave is sufficiently reduced by the poloidal coupling (e.g. $n_\|  \sim -3$) to interact with less energetic electrons, the resonant interaction is negligible. Nevertheless, the poloidal momentum of the wave changes due to the inhomogeneity of the magnetic field and the plasma density and temperature. There is a significant poloidal wave momentum gain. The wave gains poloidal momentum slowly in the non-resonant region, and then transfers it in a short distance where it resonates. However, the toroidal angular momentum of the wave does not change due to the toroidal symmetry, and the original amount is fully transferred to the plasma in the resonance region (see the constancy of the green line in Fig.1).  The non-resonant interaction can be studied as a combination of the Reynolds stress and the Lorentz force in both fluid models \cite{Klima:JPC1980,Lee:PF1983} and kinetic models \cite{Berry:PRL1999,Myra:POP2000,Jaeger:POP2000-2,Myra:POP2004}, and it has no effect on the toroidal flow \cite{Myra:POP2004, Gao:POP2007}. \\ \\
When the wave energy is transferred to the the plasma due to a resonance, the corresponding wave momentum is also transferred to the plasma. This relation has been verified by evaluating the Lorentz force in fluid models \cite{Klima:JPC1980,Elifimov:POP1994} and kinetic models \cite{Chan:PFB1993,Jaeger:POP2000,Gao:POP2006}. However, the toroidal momentum transfer by resonance has been calculated incorrectly for the LH wave \cite{Lee:RFC2011,Hellsten:PPCF2011,Wang:POP2011} resulting in an incorrect radial electric field. These calculations have ignored an important contribution to the Kennel-Engelmann quasilinear diffusion coefficient. The Kennel-Engelmann quasilinear diffusion coefficient \cite{Kennel:POF1966} describes the resonant interaction of the plasma with the wave. The gyroaverage of this quasilinear operator is used to model the diffusion of the distribution function in velocity space. However, since some components of the momentum, such as the toroidal direction, depend on the gyro-phase, the diffusion in gyro-phase must be taken into account for momentum transfer calculations, and the gyroaveraged quasilinear operator is not sufficient to explain the total toroidal momentum transfer. We reexamine the amount of momentum transfer from LH wave to the plasma by resonant interaction in this paper. The new contribution to momentum transfer that we find is important because the poloidal wave number is large in the resonance region, giving $k_\| > k_\phi$ as we have discussed above. Using the gyroaveraged quasilinear operator only transfers the parallel wave momentum, leading to an incorrect evaluation of the toroidal angular momentum transferred by the wave.\\ \\
Once the momentum of the LH wave is transferred to the electrons, part of it is transmitted to the ions by electron-ion collisions, and the rest is balanced by an electron radial pinch. The radial non-ambipolar electron pinch has been proposed as an explanation for 
the ion rotation induced by LH waves \cite{Cushman:2009,Fisch:PF1981,Helander:PPCF2005}. It has been argued that the counter-current direction momentum transfer from the LH wave to the trapped electrons induces a radially inward pinch, and it results in an additional inward radial electric field to ensure ambipolarity. This excess radial electric field will then act on the ions, leading to ion rotation \cite{Fisch:PF1981}. In this paper, we propose another (actually stronger) mechanism that gives an outward electron pinch.
\\
\\
The rest of this paper is organized as follows. In Sec. \ref{resonance}, we revisit the quasilinear diffusion operator taking into account  the flux in the gyro-phase angle direction to evaluate the total amount of momentum transfer by resonant particles. In Sec. \ref{non_resonant}, we discuss briefly the wave momentum gain or loss by non-resonant effects. The inhomogeneity of the tokamak system generates poloidal wave momentum. In Sec. \ref{Radial_particle_flux}, we discuss the radial non-ambipolar electron pinch due to the resonant momentum transfer. The pinches of the passing electrons and the trapped electrons due to the LH wave parallel momentum are explained. More importantly, we present a new outward electron pinch due to the perpendicular momentum transfer. Finally, the conclusions of this paper are given in Sec. \ref{Conclusion}.

\begin{figure} \label{npar_nphi}
\includegraphics[scale=0.6]{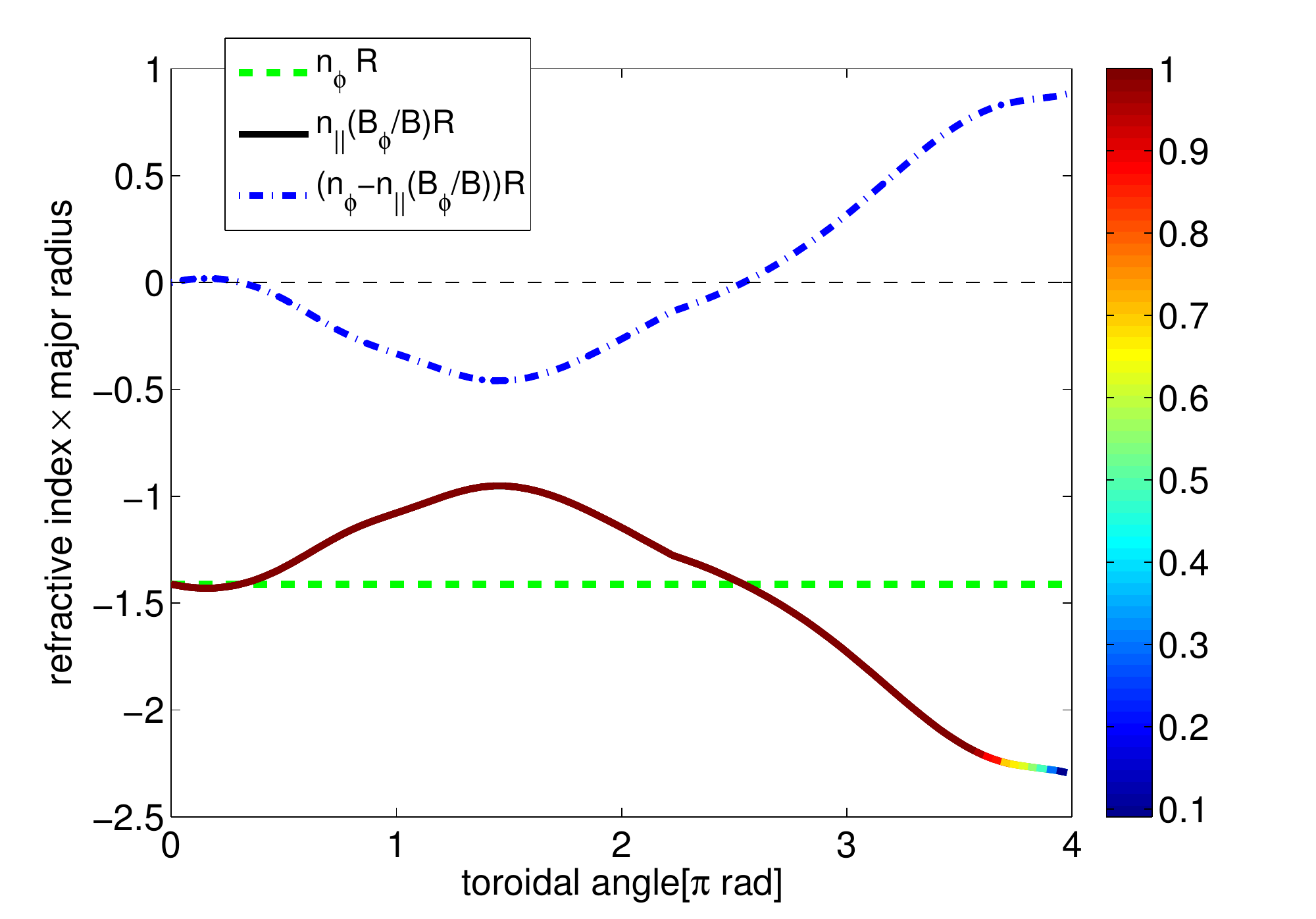}
\caption{Evolution of the toroidal angular momentum of a LH wave in terms of the propagation toroidal angle from the launcher. The solid line is the toroidal projection of the parallel refractive index multiplied by the major radius  ($n_\|( B_{\phi}/B)R$), which is an important parameter for Landau damping. The color of the solid line is the normalized Poynting flux of a ray. The power of the LH wave is absorbed by electron Landau damping beyond toroidal angle 3.5$\pi$ where its color changes from red to blue. In this power absorption region,  $n_\|( B_{\phi}/B)R$ is around -2.3, much higher than original toroidal refractive index multiplied by the major radius  ($n_\phi R \simeq -1.4 $). This graph corresponds to one of the LH wave rays for Alcator C-Mod with $B_\phi =5.3T$, plasma current $I_p=700 kA$, major radius $R_0=0.67\textrm{m}$, minor radius $a=0.22\textrm{m}$, core electron temperature $T_e=3.5KeV$, core electron density $n_e=1.2 \times 10^{20} \textrm{m}^{-3}$, initial $n_{\|} =-1.6$ and $P_{abs}=0.8MW$. These profiles are calculated using the ray tracing technique in Genray-CQL3D \cite{Harvey:1992}. The ray tracing technique can be problematic at the reflection point (toroidal angle=2.2$\pi$) where the characteristic length of the change of the plasma parameters is shorter than the wavelength. However, the upshift in $n_\|$ for strong damping (as shown beyond toroidal angle 3.5$\pi$) is widely seen in both ray tracying codes \cite{Bonoli:PF1986} and full wave codes  \cite{Wright:POP2009} for typical LH wave experimental parameters.}
\end{figure}
\clearpage
\section{Momentum transfer by resonance}\label{resonance}
\noindent In this section, we reconsider the quasilinear diffusion in velocity space including the gyro-phase. We show that we must retain the contribution to the momentum transfer from the quasilinear diffusion in the gyro-phase direction. In this section, we only focus on the momentum transfer by resonant particles and its relation to the power absorption. We derive a quasilinear diffusion operator, and we use it to evaluate the momentum deposited by the wave by applying the resonance condition. This proof can be applied to any type of resonance (cyclotron or Landau damping) and any direction of momentum. For convenience and without loss of generality, we discuss toroidal angular momentum which has both perpendicular and parallel components. Let $z$ be the direction parallel to the static magnetic field, and $x$, $y$ the orthogonal coordinates (see Figure 2-(a)). Then, using the gyro-phase angle $\alpha$, the velocity is
$\mathbf{v}=v_{\perp} \cos{\alpha} \mathbf{\hat{x}}+ v_{\perp} \sin{\alpha} \mathbf{\hat{y}}+v_{\|} \mathbf{\hat{z}}=v_{\perp} \boldsymbol{\hat{\rho}}+v_{\|} \mathbf{\hat{z}}$,
 and its toroidal component is $v_{\phi}=v_{\perp} \cos{\alpha}  ( \mathbf{\hat{x}} \cdot \boldsymbol{\hat{\phi}} )+v_{\perp} \sin{\alpha} ( \mathbf{\hat{y}} \cdot \boldsymbol{\hat{\phi}} )+v_{\|} ( \mathbf{\hat{z}} \cdot \boldsymbol{\hat{\phi}} ) $, where $ \boldsymbol{\hat{\rho}}=\mathbf{v_{\perp}}/v_{\perp}$ and $\boldsymbol{\hat{\phi}}$ is the unit vector in the toroidal direction.
 The wavenumber vector is defined as $\mathbf{k}=k_{\perp} \cos{\beta} \mathbf{\hat{x}}+ k_{\perp} \sin{\beta} \mathbf{\hat{y}}+k_{\|} \mathbf{\hat{z}}=k_{\perp} \cos{(\alpha-\beta)} \boldsymbol{\hat{\rho}}- k_{\perp} \sin{(\alpha-\beta)} \boldsymbol{\hat{\alpha}}+k_{\|} \mathbf{\hat{z}}$, and the electric field is $\mathbf{E}=E_{x} \mathbf{\hat{x}}+E_y \mathbf{\hat{y}}+E_{\|} \mathbf{\hat{z}}=E_{\perp}\boldsymbol{\hat{\rho}}+E_{\alpha}\boldsymbol{\hat{\alpha}}+E_{\|} \mathbf{\hat{z}}$, where $\boldsymbol{\hat{\alpha}}=\mathbf{\hat{z}} \times \boldsymbol{\hat{\rho}}$ is the unit vector perpendicular to both $\mathbf{v_{\perp}}$ and the magnetic field.  Here, $E_{\perp}=E_{x} \cos{\alpha}+E_{y} \sin{\alpha} = (E_{+} +E_{-})\cos{(\alpha-\beta)}-i (E_{+} -E_{-})\sin{(\alpha-\beta)}$, and $E_{\alpha}=-E_{x} \sin{\alpha}+E_{y} \cos{\alpha}= -i(E_{+} -E_{-})\cos{(\alpha-\beta)}- (E_{+} +E_{-})\sin{(\alpha-\beta)}$, where $E_{\pm}=\frac{1}{2}(E_x\pm iE_y)e^{\mp i\beta}$.\\ \\
 The quasilinear diffusion operator is obtained from
 \begin{eqnarray}
Q(f)&=&-\frac{Ze}{m}\left \langle \nabla_v \cdot \left[\left(\mathbf{E} +\frac{\mathbf{v} \times \mathbf{B} }{c}\right) {f} \right] \right \rangle \label{Q1}\\
&\simeq&-\frac{Ze}{m}\nabla_v \cdot \left[ \sum_\mathbf{k} \left \{\boldsymbol{\overline{I}} \left(1-\frac{\mathbf{k} \cdot \mathbf{v}}{\omega} \right) +\frac{\mathbf{k} \mathbf{v}}{\omega} \right\}\cdot\mathbf{E}_\mathbf{-k}   f_{\mathbf{k} } \right], 
 \label{Dql_0}
\end{eqnarray}
where the triangular bracket $\left \langle... \right \rangle $ in (\ref{Q1}) indicates the average over a number of wave periods in time and space. Here, $m$ and $Ze$ are the mass and the charge of the species of interest, respectively, $e$ is the charge of the proton, and $\boldsymbol{\overline{I}}$ is the unit tensor. We have used the Fourier analyzed perturbed fluctuating electric field, $\mathbf{E}=\sum_\mathbf{k}  \mathbf{E}_\mathbf{k}  \exp(i\mathbf{k} \cdot \mathbf{r}-i \omega_\mathbf{k} t)$, the fluctuating magnetic field $\mathbf{B}=\sum_\mathbf{k} \mathbf{B}_\mathbf{k}  \exp(i\mathbf{k} \cdot \mathbf{r}-i \omega_\mathbf{k} t)$, and the fluctuating distribution function, $f=\sum_\mathbf{k} f_\mathbf{k}  \exp(i\mathbf{k} \cdot \mathbf{r}-i \omega_\mathbf{k} t)$. The functions $\mathbf{E}_\mathbf{k} \equiv   \mathbf{E}(\omega_{\mathbf{k} },\mathbf{k})$, $\mathbf{B}_\mathbf{k} \equiv   \mathbf{B}(\omega_{\mathbf{k} },\mathbf{k})$, and $f_\mathbf{k} \equiv  f(\omega_{\mathbf{k} },\mathbf{k})$ satisfy the relation $f_{-\mathbf{k} }\equiv f(\omega_{-\mathbf{k} },-\mathbf{k})=f^*(\omega_{\mathbf{k} },\mathbf{k})$ where $*$ denotes complex conjugate and $\omega_{\mathbf{k} }=-\omega^*_{-\mathbf{k} }$. Faraday's law has been used in going from (\ref{Q1}) to (\ref{Dql_0}) to write $\mathbf{B}_\mathbf{k} =(c/\omega) \mathbf{k} \times \mathbf{E}_\mathbf{k} $\footnotemark[1]. 
\footnotetext{ In typical tokamak geometry, the toroidal and poloidal spectra are discrete due to  periodicity, but the radial spectrum is continuous. Also, the parallel spectrum is continuous for the  flux surfaces with non-rational safety factor by the coupling of toroidal and poloidal components \cite{LEE:POP2011}. Even though using integrals in Fourier space would be more appropriate, we use the notation $\sum_\mathbf{k}$ for simplicity. The summation in the discrete toroidal and poloidal spectrum space is also closer to the numerical evaluation in a code \cite{Brambilla:NF2007,Wright:IEEE2010}.}
The quasilinear operator can be written as 
 \begin{eqnarray}
Q(f)\equiv \frac{Ze}{m}\left[\frac{1}{v_{\perp} }\frac{\partial}{\partial v_{\perp}} \left(v_{\perp}\Gamma_\perp\right) +\frac{1}{v_{\perp} }\frac{\partial \Gamma_\alpha}{\partial \alpha} +\frac{\partial \Gamma_\|}{\partial v_{\|}}\right].
 \label{Dql_1}
\end{eqnarray}
The flux in the perpendicular direction is
 \begin{eqnarray}
\Gamma_\perp= -  \sum_\mathbf{k}\bigg \{E^{*}_{\mathbf{k} ,\perp}\bigg(1-\frac{k_{\|}v_{\|}}{\omega} \bigg) +E^{*}_{\mathbf{k} ,\|} \frac{k_{\perp}v_{\|}}{\omega} \cos{(\alpha-\beta)} \bigg \}f_{\mathbf{k} }
 \label{Gamma_perp},
\end{eqnarray}
the flux in the gyro-phase direction is
 \begin{eqnarray}
\Gamma_\alpha&=& -  \sum_\mathbf{k}\bigg\{E^{*}_{\mathbf{k} ,\alpha} \bigg(1-\frac{k_{\perp}v_{\perp}}{\omega} \cos{(\alpha-\beta)}-\frac{k_{\|}v_{\|}}{\omega} \bigg)\nonumber \\
&-& E^{*}_{\mathbf{k} ,\perp}\frac{k_{\perp}v_{\perp}}{\omega}  \sin{(\alpha-\beta)} -E^{*}_{\mathbf{k} ,\|} \frac{k_{\perp}v_{\|}}{\omega} \sin{(\alpha-\beta)} \bigg \}{f}_\mathbf{k} ,
 \label{Gamma_alpha}
\end{eqnarray}
and the flux in the parallel direction is
 \begin{eqnarray}
\Gamma_\|=  - \sum_\mathbf{k}\bigg\{E^{*}_{\mathbf{k} ,\|} \bigg(1-\frac{k_{\perp}v_{\perp}}{\omega} \cos{(\alpha-\beta)} \bigg) + E^{*}_{\mathbf{k} ,\perp}\frac{k_{\|}v_{\perp}}{\omega}\bigg \}{f}_\mathbf{k} .
 \label{Gamma_par}
\end{eqnarray}
Here, the perturbed fluctuating distribution function consistent with a single mode wave \cite{Stix:AIP1992} is
 \begin{eqnarray}
f_\mathbf{k}  &=& - \frac{Ze}{m}\exp({-i \mathbf{k} \cdot \mathbf{r} + i \omega t}) \int^t_{-\infty} d t^\prime \exp({i \mathbf{k} \cdot \mathbf{r}^{\prime} -i \omega t^{\prime}}) \nonumber\\
&&\times\mathbf{E_\mathbf{k} } \cdot \left[ \boldsymbol{\overline{I}} \left(1-\frac{\mathbf{v}^\prime \cdot \mathbf{k}}{\omega} \right) +\frac{\mathbf{v}^\prime \mathbf{k}}{\omega} \right] \cdot  \nabla_{v^{\prime}} f_0 ,  \label{f_k1}
\end{eqnarray}
where $(t^{\prime},\mathbf{r}^{\prime},\mathbf{v}^\prime)$ is a point of phase space along the zero-order particle trajectory. Its end point corresponds to $(t,\mathbf{r},\mathbf{v})$. The background distribution, $f_0=f_0(t,\mathbf{r},v_{\perp}, v_{\|})$, is gyro-phase independent because of the fast gyro-motion. As a result,
 \begin{eqnarray}	
f_\mathbf{k}  &=&-\frac{Ze}{m} \int_0^{\infty} d\tau \exp({i\gamma})  \bigg\{ \cos{(\eta+\Omega \tau)}(( E_{\mathbf{k} ,+}+E_{\mathbf{k} ,-})U-E_{\mathbf{k},\|}V)\nonumber\\
&&-i\sin{(\eta+\Omega \tau)}( E_{\mathbf{k} ,+}-E_{\mathbf{k} ,-})U +E_{\mathbf{k} ,\|} \frac{\partial f_0}{\partial v_{\|}}\bigg \}. \label{f_k2}
\end{eqnarray}
Here, $\tau=t-t^{\prime}$, and $\gamma=(\omega-k_{\|}v_{\|})\tau -\lambda(  \sin{(\eta+\Omega \tau)}- \sin{(\eta)}) $, where $\lambda= \frac{k_{\perp}v_{\perp}}{\Omega}$, $\eta=\alpha-\beta$,  $\Omega=ZeB_0/mc$ is the gyrofrequency and $B_0$ is the magnitude of the background magnetic field. Also, $U=\frac{\partial f_0}{\partial v_{\perp}}+ \frac{k_{\|}}{\omega}\left( v_{\perp}\frac{\partial f_0}{\partial v_{\|}}-v_{\|}\frac{\partial f_0}{\partial v_{\perp}}\right)$, and $V= \frac{k_{\perp}}{\omega}\left( v_{\perp}\frac{\partial f_0}{\partial v_{\|}}-v_{\|}\frac{\partial f_0}{\partial v_{\perp}}\right)$. We follow Stix' notation in \cite{Stix:AIP1992}.
\\
\\
For the energy transfer, the contribution of the flux in the gyro-phase direction vanishes due to the integral over $\alpha$, 
 \begin{eqnarray}
\fl \;\;\;\; P_{abs}&=&\int_{-\infty} ^{\infty} dv_{\|} \int_0^{\infty} dv_{\perp} 2\pi v_{\perp}\left \langle \frac{mv^2}{2} Q(f) \right \rangle_{\alpha}\nonumber\\
\fl &=& \int_{-\infty} ^{\infty} dv_{\|} \int_0^{\infty} dv_{\perp} 2\pi v_{\perp}\frac{Ze v^2}{2} \left[\frac{1}{v_{\perp} }\frac{\partial}{\partial v_{\perp}} \left( v_{\perp}\left \langle \Gamma_\perp\right \rangle_{\alpha}\right)+\frac{\partial\left \langle  \Gamma_\|\right \rangle_{\alpha}}{\partial v_{\|}}\right], \label{Q_avg1}
 \label{energy}
\end{eqnarray}
with $\left \langle ... \right \rangle_{\alpha}= \frac{1}{2\pi}\int_0^{2\pi} d\alpha(...)$ the gyroaverage.
For this reason, the typical Kennel-Engelmann quasilinear diffusion operator \cite{Kennel:POF1966} is gyroaveraged and does not retain the flux in the gyro-phase direction. For completeness, we have evaluated the energy transfer, $P_{abs}$, in \ref{appendixA}.
 \\
 \\
 The gyroaveraged quasilinear operator is not sufficient to calculate the toroidal momentum transfer, which has gyro-phase dependent components. The total toroidal angular momentum deposited by the wave is 
\begin{eqnarray}
P_{\phi}  &=&2\pi \int_{-\infty} ^{\infty} dv_{\|} \int_0^{\infty} dv_{\perp} v_{\perp}  \langle m R  v_{\phi}Q(f) \rangle_{\alpha} \equiv P_{\phi}^{\|} +\Delta P_{\phi}^{\perp}+\Delta P_{\phi}^{\alpha},
\end{eqnarray}
where 
\begin{eqnarray}
  P_{\phi}^{\|} &=&-ZeR \int_{-\infty} ^{\infty} dv_{\|} \int_0^{\infty} dv_{\perp} v_{\perp} \int_0^{2\pi} d\alpha ( \mathbf{\hat{z}} \cdot \boldsymbol{\hat{\phi}} )\Gamma_{\|} \label{mom_par1}
\end{eqnarray}
is the component of momentum transfer that one obtains when using the gyroaveraged quasilinear operator, whereas
\begin{eqnarray}
\fl \;\;\;\; \Delta P_{\phi}^{\perp}&=&ZeR \int_{-\infty} ^{\infty} dv_{\|} \int_0^{\infty} dv_{\perp} v_{\perp} \int_0^{2\pi} d\alpha( - \cos{\alpha}  ( \mathbf{\hat{x}} \cdot \boldsymbol{\hat{\phi}} )- \sin{\alpha} ( \mathbf{\hat{y}} \cdot \boldsymbol{\hat{\phi}}  ) )\Gamma_{\perp}\label{P_perp} \,,\\
\fl \;\;\;\;\Delta P_{\phi}^{\alpha} &=&ZeR \int_{-\infty} ^{\infty} dv_{\|} \int_0^{\infty} dv_{\perp}  v_{\perp} \int_0^{2\pi} d\alpha(  \sin{\alpha}  ( \mathbf{\hat{x}} \cdot \boldsymbol{\hat{\phi}} )- \cos{\alpha} ( \mathbf{\hat{y}} \cdot \boldsymbol{\hat{\phi}}  ) )\Gamma_{\alpha}\label{P_alpha}
\end{eqnarray}
are the contributions that appear when the complete dependence on the gyro-phase is retained. \\ \\ Using the perturbed distribution function and the expansion in Bessel functions described in \ref{appendixA}, the toroidal momentum transfer term in the parallel direction, $P^{\|}_{\phi}$, becomes
 \begin{eqnarray}
\fl P^{\|}_{\phi}&=& -\frac{{\pi} Z^2 e^2 R}{m}\sum_\mathbf{k}  \int_{-\infty} ^{\infty} dv_{\|} \int_0^{\infty} dv_{\perp} 2\pi v_{\perp}  \sum_n \delta (\omega-k_{\|}v_{\|}-n\Omega) ( \mathbf{\hat{z}} \cdot \boldsymbol{\hat{\phi}} ) \nonumber \\ \fl &\times& \frac{k_{\|}v_{\perp}^2}{\omega} |\chi_{\mathbf{k} ,n}|^2 L(f_0)\label{mom_par4} = \sum_\mathbf{k}\left\{\frac{k_\|}{\omega}P_{abs,\mathbf{k}} R( \mathbf{\hat{z}} \cdot \boldsymbol{\hat{\phi}} )\right\}= \sum_\mathbf{k}\left\{  \frac{n_\|}{c}P_{abs,\mathbf{k}} R( \mathbf{\hat{z}} \cdot \boldsymbol{\hat{\phi}} )\right\}\label{mom_par5},
\end{eqnarray}
where $\chi_{\mathbf{k} ,n}=E_{\mathbf{k} ,\|} J_{n}\frac{v_{\|}}{v_{\perp}}+  E_{\mathbf{k} ,+} J_{n-1} +E_{\mathbf{k} ,-} J_{n+1}$ is the effective electric field, and $J_{n}(\lambda)$ are the Bessel functions of the first kind  with integer order $n$. The operator $L(f_0)=\left(1-\frac{k_{\|}v_{\|}}{\omega}\right)\frac{1}{v_{\perp}}\frac{\partial f_0}{\partial v_{\perp}}+\frac{k_{\|}v_{\perp}}{\omega}\frac{1}{v_{\perp}}\frac{\partial f_0}{\partial v_{\|}}$ is introduced in \cite{Kennel:POF1966,Stix:AIP1992} (see \ref{appendixB} for the detailed derivation). The piece of the momentum transfer $P^{\|}_{\phi}$ is directly related to the quasilinear diffusion operator used to calculate the power absorption (compare equation (\ref{mom_par5}) with (\ref{P_abs2})). The direction of diffusion is determined by the characteristics of the operator $L(f_0)$ (i.e. the tangents to the contours $v_{\perp}^2+\left( v_{\|}-\frac{\omega}{k_\|}\right)^2 = $constant), and the magnitude of the diffusion is determined by the projection of the distribution function gradient onto these characteristics \cite{Kennel:POF1966,Stix:AIP1992} (see Figure 3).
In particular, for electron Landau damping of the LH wave (i.e. $\omega=k_{\|}v_{\|}$), the piece of the toroidal momentum transfer $P_{\phi}^{\|}$ can be simplified to the following equation within a small error of $O\left(\left(\frac{k_{\perp}v_{\perp}}{\Omega_e}\right)^2\frac{\Omega_e}{\omega}\right)$ :
 \begin{eqnarray} 
  \fl \;\;\;\;P_{\phi,ELD}^{\|} 
  &\simeq&- \frac{{\pi} e^2 R}{m_e}  \sum_\mathbf{k}\int_{-\infty} ^{\infty} dv_{\|} \int_0^{\infty} dv_{\perp} 2\pi v_{\perp} \delta (\omega-k_{\|}v_{\|})  |E_{\mathbf{k} ,\|}|^2 \frac{\partial f_{e0}}{\partial v_{\|}}J^2_0(\lambda) ( \mathbf{\hat{z}} \cdot \boldsymbol{\hat{\phi}} ) \nonumber \\
 \fl \;\;\;\; &=& \int_{-\infty} ^{\infty} dv_{\|} \int_0^{\infty} dv_{\perp} 2\pi v_{\perp}m_e R  v_{\phi} \frac{\partial}{\partial v_{\|}}\left( \langle \overline{D_{ql}^{ELD}} \rangle_{\alpha} \frac{\partial f_{e0}}{\partial v_{\|}}\right) \label{mom_ELD1},
\end{eqnarray}
where $m_e$ is the electron mass, and the gyroaveraged quasilinear diffusion coefficient for electron Landau damping is $\langle \overline{D_{ql}^{ELD}} \rangle_{\alpha}=\frac{\pi e^2}{m_e^2}\sum_\mathbf{k}\delta (\omega-k_{\|}v_{\|})  |E_{\mathbf{k} ,\|}|^2J^2_{0}(\lambda)$. For Landau damping, the quasilinear diffusion happens only in the parallel direction (see Figure 3). Equation (\ref{mom_ELD1}) exemplifies the problems that appear if the gyroaveraged quasilinear diffusion operator is employed to evaluate toroidal angular momentum transfer. \\ \\ Using the typical gyroaveraged quasilinear diffusion coefficient, we can only evaluate the parallel momentum transfer $P_{\phi}^{\|} $ instead of the full momentum transfer $P_{\phi}$. For the rest of the toroidal momentum transfer, we need the quasilinear diffusion operator before the gyro-phase averaging, 
 \begin{eqnarray}
\fl\Delta P_{\phi}^{\perp}+\Delta P_{\phi}^{\alpha} &=&ZeR \sum_\mathbf{k}\int_{-\infty} ^{\infty} dv_{\|} \int_0^{\infty} dv_{\perp} 2\pi v_{\perp} \int_0^{2\pi} \frac{d\alpha}{2\pi} \bigg [E^{*}_{\mathbf{k},\|} \frac{k_{\perp}v_{\|}}{\omega}\nonumber \\ \fl&\times& (\cos{\beta} ( \mathbf{\hat{x}} \cdot \boldsymbol{\hat{\phi}} )+ \sin{\beta} ( \mathbf{\hat{y}} \cdot \boldsymbol{\hat{\phi}} ))+\bigg(1-\frac{k_{\|}v_{\|}}{\omega} \bigg)(E^{*}_{\mathbf{k},x} \mathbf{\hat{x}} \cdot \boldsymbol{\hat{\phi}} +E^{*}_{\mathbf{k},y} \mathbf{\hat{y}} \cdot \boldsymbol{\hat{\phi}} ) 
 \nonumber\\\fl&&+i \bigg( \frac{k_{\perp}v_{\perp}}{\omega}\bigg) (E^{*}_{\mathbf{k},+}-E^{*}_{\mathbf{k},-})(\sin{\alpha}  ( \mathbf{\hat{x}} \cdot \boldsymbol{\hat{\phi}} )- \cos{\alpha} ( \mathbf{\hat{y}} \cdot \boldsymbol{\hat{\phi}}  ))  \bigg] f_{\mathbf{k}}.\label{mom_perp0}
\end{eqnarray}
Using \ref{appendixB}, we can simplify this equation to     
\begin{eqnarray}
\fl&\Delta P_{\phi}^{\perp}+\Delta P_{\phi}^{\alpha} =
-\frac{{\pi} Z^2e^2 R}{m} \sum_\mathbf{k} \int_{-\infty} ^{\infty} dv_{\|} \int_0^{\infty} dv_{\perp} 2\pi v_{\perp}   \nonumber\\ \fl& \times \sum_n \delta (\omega-k_{\|}v_{\|}-n\Omega)  (\cos{\beta} ( \mathbf{\hat{x}} \cdot \boldsymbol{\hat{\phi}} )+ \sin{\beta} ( \mathbf{\hat{y}} \cdot \boldsymbol{\hat{\phi}} )) \frac{k_{\perp}v_{\perp}^2}{\omega} |\chi_{\mathbf{k} ,n}|^2 L(f_0)\label{mom_perp1}\nonumber
\\\fl&=\sum_\mathbf{k}\left( \frac{\mathbf{k_{\perp}} \cdot {\boldsymbol{\hat{\phi}}}}{\omega}P_{abs,_\mathbf{k}} R \right )= \sum_\mathbf{k}\left\{ \frac{n_\perp}{c}P_{abs,_\mathbf{k}} R  (\cos{\beta} ( \mathbf{\hat{x}} \cdot \boldsymbol{\hat{\phi}} )+ \sin{\beta} ( \mathbf{\hat{y}} \cdot \boldsymbol{\hat{\phi}} ))\right\}
\label{P_add5}.
  \end{eqnarray}
  \\
  The perpendicular momentum transfer, $\Delta P_{\phi}^{\perp}+\Delta P_{\phi}^{\alpha}$, cannot change the gyro-averaged distribution function as shown in Fig. 3. As a result, it cannot drive a parallel current, while a perpendicular energy transfer (e.g. in the electron cyclotron current drive (ECCD) \cite{Fisch:RMP1987}) can drive the parallel current through collisions because it can change the gyro-averaged distribution function in the perpendicular direction.
\\  
\\In conclusion, for any resonance (e.g. cyclotron, Landau damping), the total toroidal angular momentum transfer according to (\ref{mom_par5}) and (\ref{P_add5}) is
  \begin{eqnarray}
P_{\phi}  = P_{\phi}^{\|} +\Delta P_{\phi}^{\perp}+\Delta P_{\phi}^{\alpha}=\sum_\mathbf{k}\left (\frac{\mathbf{k} \cdot {\boldsymbol{\hat{\phi}}}}{\omega}P_{abs,\mathbf{k}} R \right),\label{mom_phi}
\end{eqnarray}
as expected \cite{Bers:1972,Klima:JPC1980,Elifimov:POP1994,Chan:PFB1993,Jaeger:POP2000,Gao:POP2006}. The toroidal angular momentum absorbed in the plasma is equal to the launched momentum only when both the parallel and the perpendicular momentum are taken into account correctly, as shown in Fig. 2-(b). If only the parallel momentum transfer by resonant interaction is considered as is done in \cite{Lee:RFC2011,Hellsten:PPCF2011,Wang:POP2011}, it gives the incorrect result that the toroidal momentum transfer is larger than the launched toroidal momentum. 
\\ \\

\begin{figure}
\includegraphics[scale=0.5]{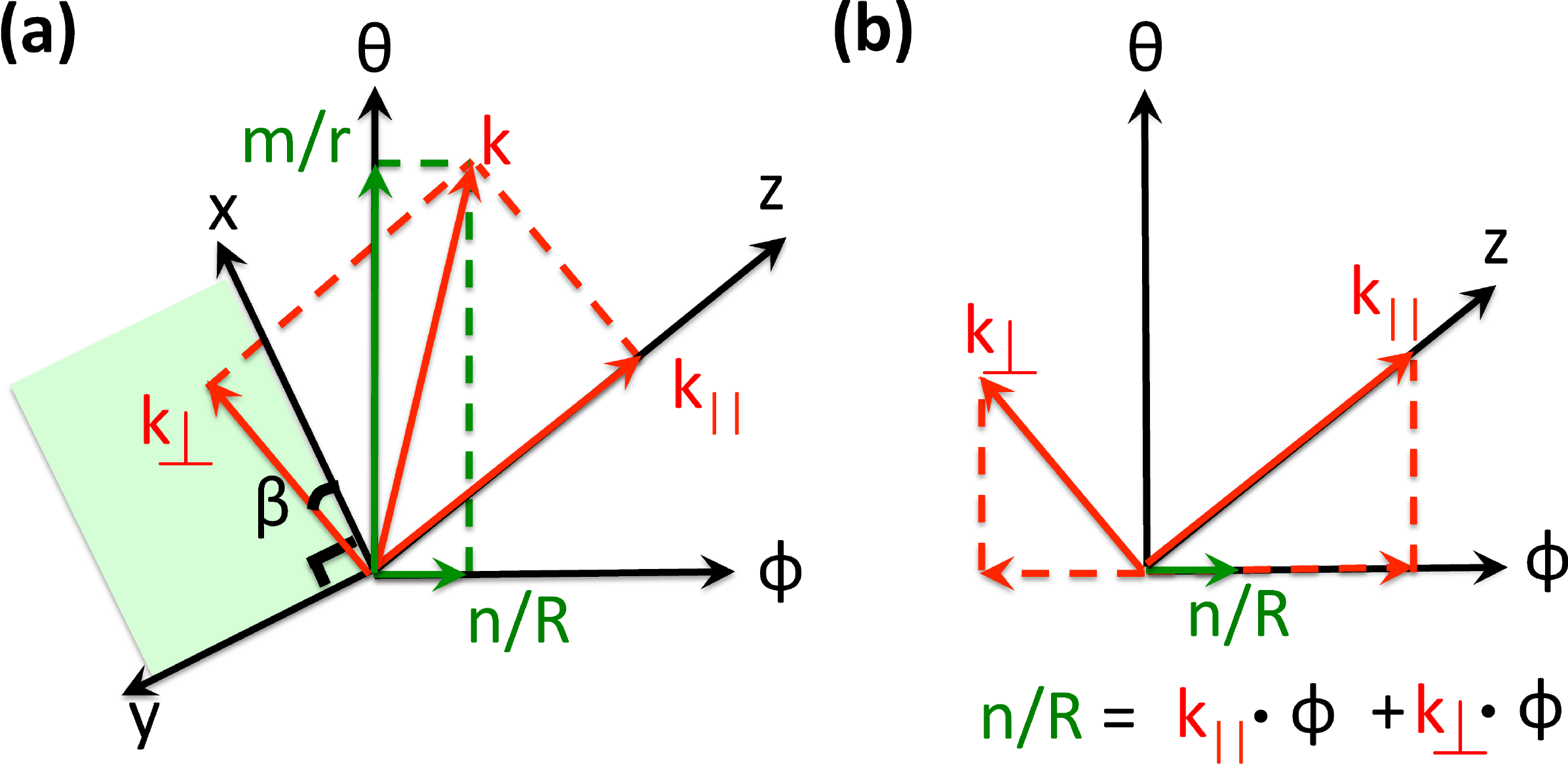}
\caption{
(a) Sketch of a wave vector $\mathbf{k}=(\textrm{m}/r) \boldsymbol{\hat{\theta}}+(\textrm{n}/R) \boldsymbol{\hat{\phi}}=\mathbf{k_{\perp}}+\mathbf{k_\|}$ in a parallel ($z$)-perpendicular ($x,y$) coordinate system and in a toroidal ($\phi$)-poloidal ($\theta$) coordinate system. Here, $\textrm{n}$ and  $\textrm{m}$ are the toroidal and poloidal wave number, respectively, and the radial wave vector is not represented, because it cannot contribute to the toroidal momentum. For the LH wave, the component $k_{\|}$ has a bigger toroidal projection than the initial toroidal component $\textrm{n}/R$ at the launcher due to the poloidal coupling. (b) Sketch of the toroidal momentum conservation.  The sum of the toroidal projection of  $k_{\|}$ and $k_{\perp}$ is equal to the launched toroidal wave vector $\textrm{n}/R$.} 
 \end{figure}
 \begin{figure}
\includegraphics[scale=0.6]{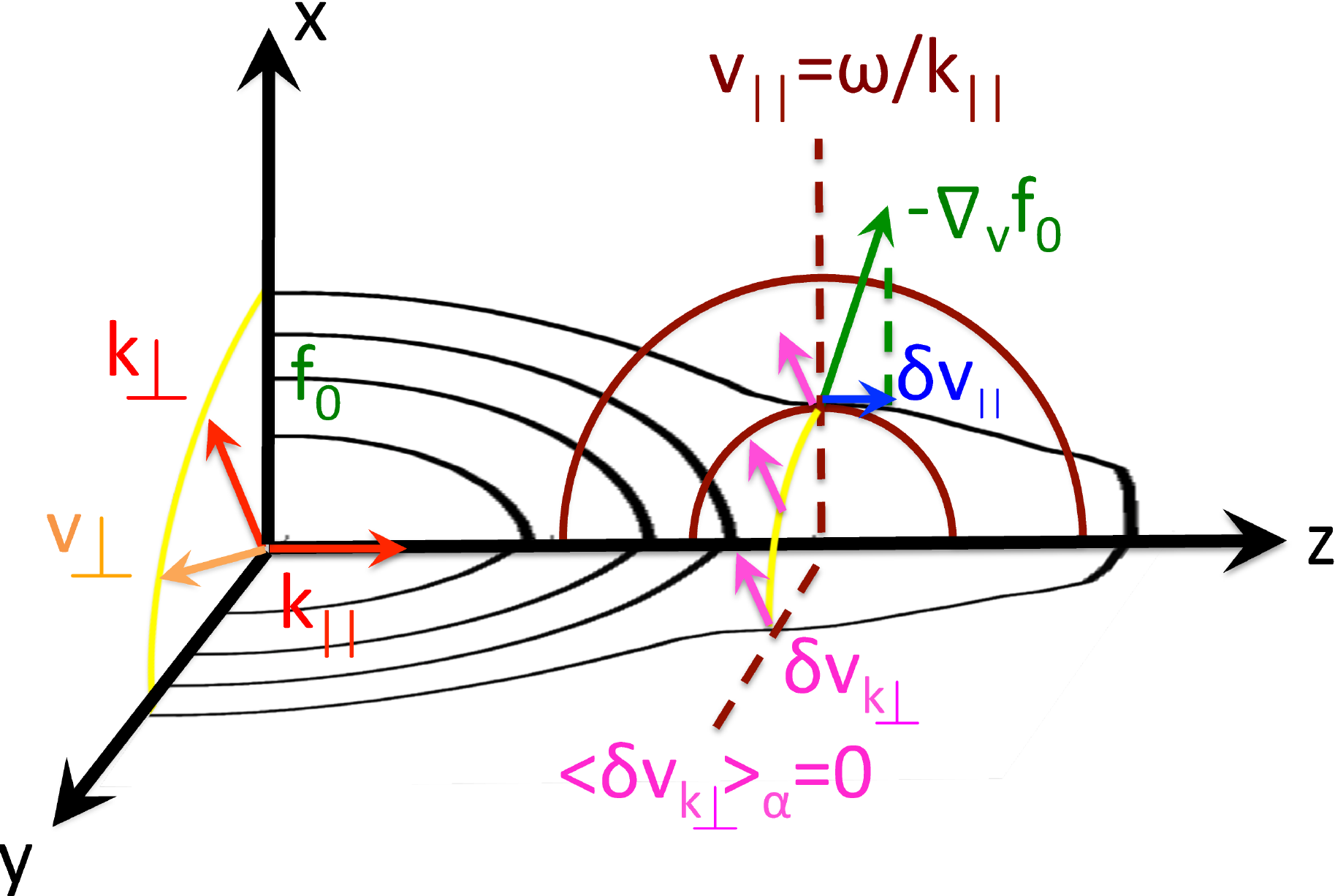}
\caption{
Sketch of the quasilinear diffusion direction and magnitude in $v_{\|}-v_{\perp}$ space (i.e. a parallel ($z$)-perpendicular ($x,y$) coordinate system). The black contours in the $x-z$ plane and the $y-z$ plane are the contours of the gyroaveraged distribution function and the brown contours are the characteristics of the operator $L(f_0)$. The diffusion direction is tangential to the characteristics, $v_{\perp}^2+\left( v_{\|}-\frac{\omega}{k_\|}\right)^2 = $constant. For Landau damping, the intensity of the diffusion is determined by the projection of $\nabla_v f_0$ onto the characteristics of the operator $L(f_0)=0$ at $v_{\|}=\omega/ k_\|$. The diffusion in velocity space results in an average increase of the parallel velocity, $\delta v_\|$. The perpendicular momentum transfer $\delta v_{k_\perp}$ has the direction of $k_{\perp}$, but its effect on the distribution function vanishes due to the fast gyro-motion (the averaged perpendicular acceleration represented by the pink arrows in the yellow circle in the x-y plane vanishes).}
 \end{figure}
 
\clearpage
\section{Momentum transfer by non-resonant interaction}\label{non_resonant}
The increase in the poloidal mode number is important to determine the location of the resonance as shown in Fig.1. In the eikonal limit, the poloidal mode number of the LH wave ($\textrm{m}\simeq k_\theta r$ in a circular tokamak) is determined by the poloidal variation of the determinant of the dispersion relation $D_0$ along a ray path \cite{Bonoli:PF1986},
\begin{eqnarray}
\frac{d\textrm{m}}{dt}&=&\frac{\partial D_0(\omega, \textrm{m}, n_e, T_e, B_\phi)}{\partial \theta} \bigg/ \frac{\partial D_0(\omega, \textrm{m}, n_e, T_e, B_\phi)}{\partial \omega}.\label{dmdtheta} 
\end{eqnarray}
Here, in the electrostatic limit, $D_0 \simeq S k_{\perp}^2+P k_{\|}^2$, $S \simeq 1+ \omega_{pe}^2/{\Omega_e^2}- \omega_{pi}^2/{\omega^2} \sim O(1)$ and $P  \simeq -\omega_{pe}^2/{\omega^2} \gg S$ are the components of the dielectric tensor, and $\omega_{pe}$ and  $\omega_{pi}$ are electron and ion plasma frequency, respectively. Figure 1 shows that as the LH wave propagates from the low field side launcher in an inhomogeneous tokamak, the increase in $\textrm{m}$ can be as large as $\textrm{n}q$ at the resonance position. Here, $\textrm{n} \simeq k_\phi R$ and $q \simeq \frac{r}{R}\frac{B_\phi}{B_\theta}$ are the toroidal mode number and the safety factor, respectively.
\\
\\
 As we discussed in Section \ref{resonance}, the poloidal wave momentum is transferred to the electrons mainly by resonance, but there is another mechanism that makes the poloidal number much larger at the resonance than at the launcher. 
The origin of the increased poloidal momentum is the external force required to keep the density $n_e$, the temperatures $T_e$ and $T_i$, and the static magnetic field $B_\phi$ constant in time in the dispersion relation. We assume that these parameters are fixed in the dispersion relation because the transport and the resistive time scale are much longer than the propagation time of the wave. The wave exerts a non-resonant force that can affect the evolution of the background profile. In general, this non-resonant force is smaller than the resonant one by a factor of $v_g/(\gamma L) <1$, where L is the characteristic length of variation of the background, $v_g$ is the group velocity of the wave, and $\gamma$ is the wave damping rate at the resonance region. However, the accumulated momentum transfer by the non-resonant force along the ray path is not negligible.\\ \\
The nonlinear forces due to the RF wave have been investigated in previous works \cite{Klima:JPC1980,Lee:PF1983,Berry:PRL1999,Myra:POP2000,Jaeger:POP2000-2,Myra:POP2004}. For example, the nonlinear force exerted by the wave has been calculated for a tokamak by neglecting the gradient of the magnetic field compared to the gradient of the density and the temperature \cite{Myra:POP2000}. In a steady state, these nonlinear forces must exactly balance the momentum increase of the wave. Thus, the wave takes momentum from the plasma as it propagates due to the plasma inhomogeneity, leading to the increase in the wave poloidal momentum. This momentum is given back to the plasma at the resonance position. Consequently, the wave has redistributed the poloidal momentum of the plasma. The effect that this has on the poloidal rotation is small due to the strong poloidal collisional damping in a tokamak \cite{Hirshman:NF1981}. 

\section{Radial particle flux by LH wave}\label{Radial_particle_flux}
In this section, we investigate different types of electron radial drifts that can be induced by the resonant momentum transfer from the LH wave. The Lorentz force due to the radial electron pinch, and the collisional friction in the parallel direction can balance the toroidal force due to waves. The dominant radial electron drift comes from toroidal momentum transfer in the perpendicular plane  (Sec. \ref{e_transport1}).  The Lorentz force that results from the pinch is comparable ($O(100\%) $) to the LH wave momentum source, giving a sizeable radial pinch (O(1mm/sec)) that has an outward direction in tokamaks. Other radial pinches induced by the wave parallel momentum transfer are relatively small. The passing electron pinch caused by the resonance gives a Lorentz force which is $O(10\%) $ of the LH wave momentum (Sec. \ref{e_transport2}), and the Ware-like LH wave induced pinch by trapped electrons \cite{Fisch:PF1981} is associated with only $O(1\%)$ of the LH wave momentum transfer (Sec. \ref{e_transport3}).
\subsection{Outward electron pinch due to perpendicular wave momentum}\label{e_transport1}
The quasilinear term due to the LH wave in the Fokker-Plank equation gives rise to a correction to the electron distribution function, $ F_e=f_e^{\prime}-f_e$, where $f_e^{\prime}$ and $f_e$ are the electron distribution function with and without LH wave respectively. For convenience, we write $F_e$ as a function of total energy $E=\frac{1}{2}m_e v^2-e\Phi$, where $\Phi$ is the background potential, magnetic moment $\mu=\frac{m_ev_\perp^2}{2B}$, and the gyro-phase angle $\alpha$. The equation for $F_e$ in these variables is
\begin{eqnarray}
\fl \;\;\;\; \frac{\partial F_e}{\partial t}-e\frac{\partial \Phi}{\partial t}\frac{\partial F_e}{\partial E}+v_{\|}\nabla_{\|}F_{e} +  \mathbf{v_d} \cdot \nabla F_e+\Omega_e \frac{\partial F_{e}}{\partial \alpha}=C_e(F_{e})+ Q_{LH}(f_e^{\prime})\label{Fokker3},
\end{eqnarray}
where  $C_e(f_e)$ is the linearized collision operator to the order of interest (i.e. $C_e(F_e)=C_{ee}(F_e,f_e)+C_{ee}(f_e,F_e)+\sum_i C_{ei}(F_e,f_i)$), and $\mathbf{v_d}$ is the $\nabla B$ and curvature drift. In (\ref{Fokker3}) we only consider the long wavelength and slowly evolving piece of the distribution function because the quasilinear term affects mainly the background distribution function. The size of the first and second terms is determined by the gyro-Bohm transport time scale, $\partial /\partial t \sim D_{gB}/a^2 \sim \epsilon_i^2 v_{ti}/a \sim \sqrt{m_i/m_e} \epsilon_e^2 v_{te}/a$, making it much smaller than other terms in (\ref{Fokker3}). Here, $D_{gB}=\epsilon_i \rho_i v_i$ is the gyro-Bohm diffusion coefficient, $a$ is the minor radius, $m_i$ is the ion mass, $v_{te}$ and $v_{ti}$ are the electron and ion thermal velocities, and  $\epsilon_e=\rho_e/a \ll 1$, $\epsilon_i=\rho_i/a\sim \sqrt{m_i/m_e}\epsilon_e \ll 1$ are the small ratios of electron and ion Larmor radius over the radial scale length, respectively. The third term in  (\ref{Fokker3}) is of order $v_{te}F_e/(qR)$ where $q$ is the safety factor and $R$ is the major radius. The fourth term in (\ref{Fokker3}) is smaller than the third term by $(B/B_{\theta}) \epsilon_e$. The gyro-motion term $\Omega_e \frac{\partial F_{e}}{\partial \alpha}$ is much larger than any of the other terms (i.e.  $v_{\|}\nabla_{\|}F_{e} /\left(\Omega_e \frac{\partial F_{e}}{\partial \alpha}\right)\sim a\epsilon_e/(qR) \ll 1$  and  $ C(F_{e}) /\left(\Omega_e \frac{\partial F_{e}}{\partial \alpha}\right)\sim \nu_e/\Omega_e  \ll 1$).  Then the lowest order equation is trivial,  $\Omega_e \frac{\partial {F_{e0}}}{\partial \alpha}=0$ (i.e. $F_{e0}= \left \langle F_{e}\right\rangle_{\alpha}$), and the next order equation is 
\begin{eqnarray}
v_{\|}\nabla_{\|}F_{e0} +  \Omega_e \frac{\partial F_{e1}}{\partial \alpha}=C_e(F_{e0})+ Q_{LH}(f_e^{\prime})\label{epinch0}.
\end{eqnarray}
Here, we have neglected the time derivative term and the perpendicular drift term. 
The gyro-phase independent part can be obtained by taking the gyro-average of (\ref{epinch0}),
\begin{eqnarray}
v_{\|}\nabla_{\|}F_{e0}=C_e( F_{e0})+ \left \langle Q_{LH}(f_e^{\prime})\right\rangle_{\alpha} .\label{DKE_Q2}
\end{eqnarray}
The quasilinear term balances with the collision operator and the parallel streaming term. The gyro-phase dependent part, $\widetilde{F_{e}}=F_e-\left \langle F_{e}\right\rangle_{\alpha}$, is obtained from the gyro-phase dependent contribution to equation (\ref{epinch0}), giving
\begin{eqnarray}
\fl \;\;\;\; \Omega_e \frac{\partial \widetilde{F_{e1}}}{\partial \alpha}&=&Q_{LH}(f_e^{\prime})- \left \langle Q_{LH}(f_e^{\prime})\right\rangle_{\alpha}\nonumber \\\fl \;\;\;\; &=&-\frac{e}{m_e}\left[\frac{1}{v_{\perp} }\frac{\partial}{\partial v_{\perp}} v_{\perp}(\Gamma_\perp-\left \langle \Gamma_\perp \right\rangle_{\alpha}) + \frac{1}{v_{\perp} }\frac{\partial\Gamma_\alpha}{\partial \alpha} +\frac{\partial}{\partial v_{\|}}(\Gamma_\|-\left \langle \Gamma_\| \right\rangle_{\alpha}) \right]\label{DKE_Q40}.
\end{eqnarray}
Its solution is
\begin{eqnarray}
\fl \;\;\;\; \widetilde{F_{e1}}&=&-\frac{e}{m_e\Omega_e} \int d\alpha \left[\frac{1}{v_{\perp} }\frac{\partial}{\partial v_{\perp}} v_{\perp}(\Gamma_\perp-\left \langle \Gamma_\perp \right\rangle_{\alpha}) +\frac{1}{v_{\perp} }\frac{\partial \Gamma_\alpha}{\partial \alpha} +\frac{\partial}{\partial v_{\|}}(\Gamma_\|-\left \langle \Gamma_\| \right\rangle_{\alpha})\right]\nonumber \\
\fl \;\;\;\; &\sim& \frac{ \Delta P_{\phi}^{\perp}+\Delta P_{\phi}^{\alpha}}{n_e m_e v_{te}R\Omega_e}f_{Me}\label{DKE_Q4}.
\end{eqnarray}
Thus, the collisional toroidal friction due to the gyro-phase dependent piece of the distribution function is much smaller than the corresponding RF force, 
\begin{eqnarray}
\int dv^3 (m_e R v_{\phi})C(\widetilde{F_{e1}})\sim \left(\frac{\nu_e}{\Omega_e}\right)\left( \Delta P_{\phi}^{\perp}+\Delta P_{\phi}^{\alpha}\right)\label{DKE_Q5},
\end{eqnarray}
and most of the perpendicular momentum transfer is balanced by the Lorentz force $\int dv^3 (m_e R v_{\phi})\Omega_e \frac{\partial \widetilde{F_{e1}}}{\partial \alpha}$ from (\ref{DKE_Q40}). \\ \\
The radial particle flux can be obtained from
\begin{eqnarray}
\fl \left \langle \mathbf{\Gamma_e} \cdot \nabla \psi \right \rangle_{s}
\simeq \left \langle \int dv^3 F_{e0} \mathbf{v_d} \cdot  \nabla \psi \right \rangle_{s}+ \left \langle \int dv^3 \widetilde{F_{e1}} v_{\perp} \boldsymbol{\hat{\rho}} \cdot  \nabla \psi \right \rangle_{s}\label{gamma_e1},
\end{eqnarray} 
where $\psi$ is the poloidal magnetic flux, $\left \langle ... \right \rangle_{s}$ is the flux-surface average (see \ref{appendixC}) and $\mathbf{\Gamma_e}$ is the electron particle flux due to the correction $F_e$.
From the steady state Fokker-Planck equation given in (\ref{epinch0}), 
taking the moment ($m_e v_{\phi} R $) and a flux-surface average of the resulting moment equation, we can relate the radial pinch $\left \langle \mathbf{\Gamma_e}  \cdot \nabla \psi \right \rangle_{s}$ to the correction $F_e$ by
\begin{eqnarray}
\fl \frac{e}{c} \left \langle \mathbf{\Gamma_e} \cdot \nabla \psi \right \rangle_{s}
&\simeq&\left \langle \frac{n_\phi-n_\|( \mathbf{\hat{z}} \cdot \boldsymbol{\hat{\phi}} )}{c}RP_{abs}\right \rangle_{s}\nonumber\\ &+&\left \langle(\mathbf{\hat{z}} \cdot \boldsymbol{\hat{\phi}}) R\int d^3v m_e v_{\|} \left [C( F_{e0})+\left \langle Q_{LH}(f_{e}^{\prime})\right\rangle_{\alpha}\right ] \right \rangle_{s} \label{epinch1}.
\end{eqnarray} 
To obtain equation (\ref{epinch1}), we use that the first and second term on the left hand side of (\ref{epinch0}) give the first and second term on the right hand side of (\ref{gamma_e1}), respectively. The right hand side of (\ref{epinch1}) is obtained by decomposing the right hand side of  (\ref{epinch0}) into the gyro-phase dependent and gyro-phase independent pieces,
\begin{eqnarray}
 \fl \;\;\;\;C_e(F_{e})+ Q_{LH}(f_e^{\prime}) \simeq \left [Q_{LH}(f_e^{\prime})- \left \langle Q_{LH}(f_e^{\prime})\right\rangle_{\alpha}\right]+\left [C_e( F_{e0})+\left \langle Q_{LH}(f_{e}^{\prime})\right\rangle_{\alpha}\right ]\label{epinch_r0}.
\end{eqnarray}
The second term on the right hand side of (\ref{epinch1}) is the parallel force balance obtained from the second term on the right hand side of (\ref{epinch_r0}), which will be discussed in the next subsection. The first term on the right hand side of (\ref{epinch1}) is the toroidal projection of the perpendicular wave momentum transfer, $\left( \Delta P_{\phi}^{\perp}+\Delta P_{\phi}^{\alpha}\right)$, which comes from the first term on the right hand side of (\ref{epinch_r0}). In (\ref{epinch_r0}) we have already neglected the perpendicular collisional friction (see (\ref{DKE_Q5})).\\ \\ 
The collisions transfer most of the parallel wave momentum to the ions, but the rest of the toroidal angular momentum (e.g. $n_\phi R - n_\|( B_{\phi}/B)R \simeq 0.9$ in Fig. 1) remains and it has the opposite toroidal direction to the original toroidal angular wave momentum, giving an electron outward pinch that is opposite to the inward pinch predicted in previous works \cite{Cushman:2009,Fisch:PF1981,Helander:PPCF2005}. Physically, the outward radial pinch comes from the effect  of the perpendicular wave momentum transfer $\Delta P_{\phi}^{\perp}+\Delta P_{\phi}^{\alpha}$ on the gyro-motion (see Fig. 4-(a)). This electron pinch is still very small compared to the Ware pinch \cite{Ware:PRL1970}. For example, if $1MW$ of LH wave power is locally absorbed in a volume of $0.1$ m$^3$ where the plasma density is $10^{20} $m$^{-3}$, the poloidal magnetic field is $B_{\theta}=0.5T$, and the refractive index is $n_\phi  - n_\|( B_{\phi}/B) =1$, then the electron outward radial pinch is about $4$ mm/s which is a hundred times smaller than the Ware pinch for a DC toroidal electric field of 0.2 V/m. 
\\
\\
We assume that the electron transport time scale in equation (\ref{Fokker3}) is much longer than the time scale of the LH wave momentum and energy transfer. The LH wave momentum transfer is balanced by collisions and the Lorentz force due to the radial electron pinch in a short time (O(10 $\mu$sec)-O(1 msec)). The new outward radial particle pinch in this paper does not cause a significant radial transport of the toroidal momentum, because it is typically smaller than the turbulent particle pinch. Instead, the outward radial electron pinch only transfers the toroidal momentum from the electrons to the ions at the local flux surface because the ions follows the radial motion of the electrons due to the ambipolarity condition. The momentum transfer by the radial particle pinch happens in an ion transit time scale (O(10 $\mu$sec)-O(1 msec)) in which the ion classical and neoclassical polarization can respond to the electron radial current \cite{Hilton:PPCF1999}. The initial direction of toroidal momentum that the ions gain from the LH wave is determined by the transfer mechanism having the shorter time scale among the ion-electron collisions that transfer parallel momentum, and the radial outward ion pinch that transfers perpendicular momentum. The comparison between these time scales determines the initial direction because the parallel and perpendicular momentum transfers typically have the opposite signs of toroidal momentum as shown in Figure 2-(b). However, as soon as both parallel and perpendicular momentum are transferred to the ions, the ions achieve the original size and direction of the launched LH wave toroidal angular momentum. Then, in the ion transport time scale (O(1 msec)-O(100 msec)), due to the turbulent transport of the ion toroidal momentum (a turbulent viscosity), the momentum is radially transferred out \cite{Parra:PPCF2010}. Eventually, the change of ion toroidal rotation by LH wave is saturated and the system (a tokamak) can reach steady state : The input from the LH wave balances the output due to the ion momentum turbulent transport.

\subsection{Passing electron pinch due to parallel wave momentum}\label{e_transport2}
To solve for the gyro-phase independent perturbation $\left \langle F_{e}\right\rangle_{\alpha}$ due to the LH wave in (\ref{DKE_Q2}), we use a subsidiary expansion of $F_{e0}=F_{e0}^0+F_{e0}^1+...$, in the small ratio of the collision frequency over the transit frequency in the banana regime. The lowest order equation is $\nabla_{\|}F_{e0}^0=0$,  implying that $F_{e0}^0$ is a flux function. The next order equation is
\begin{eqnarray}
v_{\|}\nabla_{\|}F_{e0}^1=C_e( F_{e0}^0)+ \left \langle Q_{LH}(f_e^{\prime})\right\rangle_{\alpha}.\label{DKE_Q20}
\end{eqnarray}
Taking a bounce average of (\ref{DKE_Q20}) (see \ref{appendixC}), the left hand side of (\ref{DKE_Q20}) vanishes,
\begin{eqnarray}
\left \langle C_e(F_{e0}^0) + \left \langle Q_{LH}(f^{\prime}_{e})\right\rangle_{\alpha} \right \rangle_{b}=0 \label{bouncing0}.
\end{eqnarray}
According to (\ref{b_vs_s1}), for passing particles, this equation is equivalent to
\begin{eqnarray}
\left \langle \frac{B}{v_{\|}}\left(C_e(F_{e0}^0)+ \left \langle Q_{LH}(f_{e}^{\prime})\right\rangle_{\alpha}\right) \right \rangle_{s}=0.   \label{bouncing1}
\end{eqnarray}
In general, the solution $F_{e0}^0$ to equation (\ref{bouncing1}) does not make the second term on the right hand side of (\ref{epinch1}) vanish, giving a non-zero radial pinch due to the passing electrons. This imbalance comes from the variation of $v_\|$, $B$ and $R$ along the orbit, which is of the order of the local aspect ratio, $O(r/R)$. For the electrons resonant with the LH wave, the effect of the change of $v_\|$ along the orbit is negligible because most resonant electrons have much larger parallel velocity than perpendicular velocity (i.e. small magnetic moment, $\mu \simeq 0$) due to the high phase velocity of the wave in the parallel direction. The non-vanishing contribution to the radial pinch is due the competition between the localized wave power absorption within a flux surface and the collisions that occur over the whole flux surface.\\ \\
Physically, this pinch can be explained by how the passing orbit of a single electron is changed by the resonance. The canonical angular momentum of the electron $\psi^*=\psi+I v_\|/\Omega_e$ determines the radial deviation of the electron orbit from the flux surface $\psi$ due to the curvature and $\nabla B$ drifts. Here $I=RB_{\phi}$ is a flux function to lowest order. After the resonance with the negative $k_\|$ of the LH wave, the absolute value of the negative velocity of the resonant electrons is increased by  $|\Delta v_\||$ due to the absorbed wave power. Accordingly, the change of the canonical momentum is $\Delta \psi^*=I \Delta v_\|/\Omega_e <0$, where the gyrofrequency $\Omega_e$ is evaluated at the local resonance point within the flux surface. Assuming the low frequency collisions cause the resonant electron to lose its momentum only after many transits, we can use the temporally averaged radial location to describe its radial motion. The increase in the transit averaged (bounce averaged) radial position of the particle is 
\begin{eqnarray}
\Delta  \left \langle \psi\right \rangle_{b}&=&\left \langle \psi_2 \right \rangle_{b}-\left \langle \psi_1 \right \rangle_{b}= \Delta \psi^*- \left\{ \left \langle \left(\frac{I  v_\|}{\Omega_e}\right)_2 \right \rangle_{b}-\left \langle \left(\frac{I  v_\|}{\Omega_e}\right)_1 \right \rangle_{b}\right\} \label{single0} \\
&=& \frac{I\Delta v_\|}{\Omega_e} - \frac{m_e c I }{e}\left( \frac{\oint_2 dl/B}{\tau_{b2}}-\frac{\oint_1 dl/B}{\tau_{b1}}\right) \label{single1} \\&\simeq& I \Delta v_\| \left(\frac{1}{\Omega_e} - \frac{1}{\Omega_{e0}} \right).  \label{single2} 
\end{eqnarray}
Here, the values with the subscripts 1 and 2 are before and after the resonance, respectively, and $\tau_b$ is the bounce time. From (\ref{single0}) to (\ref{single1}), equation (\ref{b_vs_s1}) is used. From (\ref{single1}) to (\ref{single2}), we neglect the radial displacement due to the poloidal variation of the parallel velocity because of the small magnetic moment ($\mu \simeq 0$). The flux surface averaged value in (\ref{single1}) is approximated by that at the magnetic axis using a small inverse aspect ratio expansion. The frequency $\Omega_{e0}$ is the gyrofrequency at the magnetic axis. Equation (\ref{single2}) means that the temporally averaged particle radial flux due to the resonance is negative for a low field side resonance (inward radial pinch) and positive for a high field side resonance (outward radial pinch). It is shown in Fig. 4-(b). The increase in the curvature drift due to the increase in the parallel velocity after the resonance results in the different passing orbits depending on the resonance location on the flux surface. This radial drift is included in the second term on the right hand side of (\ref{epinch1}) as the competition between the localized wave power absorption within a flux surface and the collisions that occur over the whole flux surface.
For a typical small inverse aspect ratio tokamak, this imbalance is small, about 10$\%$ (O(r/R)) of the total momentum transfer. \\ 
\subsection{Trapped electron pinch due to parallel wave momentum}\label{e_transport3}
For trapped electrons, since odd functions in $v_{\|}$ vanish under the bounce averaging according to (\ref{b_vs_s2}), equation (\ref{bouncing0}) becomes
\begin{eqnarray}
\left \langle C_e(F^0_{e0})\right \rangle_{b}= -\left \langle\left \langle Q_{LH}^{even}(f_{e}^{\prime})\right\rangle_{\alpha}\right \rangle_{b}  \label{bouncing2}.
\end{eqnarray}
The trapped particle contribution to the distribution function $F_{e0}$ is an even function of $v_\|$, because the bounce averaged quasilinear term is even. Two trapped electrons at the outer-midplane that have opposite parallel velocities have the same electron Landau damping resonance at the same local point in their banana orbits due to the small electron banana width (see Fig. 4-(c)). The non-zero $F_{e0}$ due to the LH wave can be understood as follows: a trapped electron is accelerated only when it resonates with the wave, that is, when its velocity is the same as the wave phase velocity, and this acceleration continues every transit until it collides. As a result, it has an open trajectory that moves inward every bounce. There is no net gain of toroidal angular momentum for the trapped electron because $F_{e0}$ is an even function, but there is a gain of canonical angular momentum that leads to a Ware-like LH induced pinch. This pinch is the contribution of the trapped electrons to the second term on the right hand side of (\ref{epinch1}). The LH wave trapped electron pinch is tiny, because the power absorption by trapped electron is less than $1\%$ of total wave power due to the small size of the population of trapped electrons resonant with the LH wave phase velocity ($\omega/k_{\|} \sim 3v_{th}-10v_{th}$) . It results in a very small contribution to the radial pinch (approximately less than 0.1mm/s). \\ \\
The mechanism behind the radial pinch by trapped electrons is similar to the mechanism of the outward radial pinch due to the gyro-motion described in section \ref{e_transport1}. Instead of considering the effect of the acceleration on the gyro-motion, one needs to consider its effect on the banana orbit (compare Fig. 4-(a) and Fig 4-(c)).  
\begin{figure} \label{banana_orbits}
\includegraphics[scale=0.6]{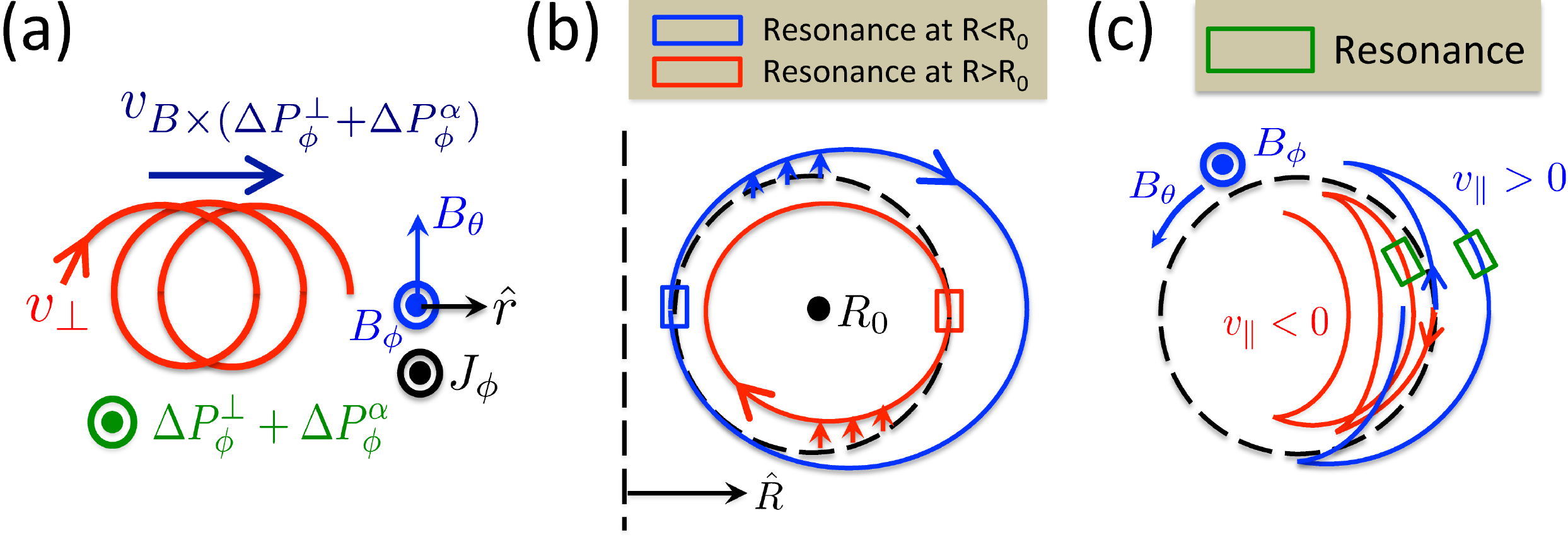}
\caption{(a) Sketch of the outward radial drift during the gyro-motion due to the perpendicular momentum transfer, $\left( \Delta P_{\phi}^{\perp}+\Delta P_{\phi}^{\alpha}\right)$. The direction of the toroidal component of the perpendicular force is the same as the direction of the plasma current $J_\phi$. (b) Sketch of the poloidal cross section of the trajectories for two passing electrons that receive wave parallel momentum at the inner-midplane (blue) and the outer-midplane (red), respectively. The dashed black line is the transit orbit before the resonances. Due to the increased curvature drift (upward direction) after the resonance, the passing electron orbits are different depending on the resonance location. The temporally averaged radial flux is outward for a resonance at $R<R_0$ and inward for a resonance at  $R>R_0$. (c) Sketch of the poloidal cross section of the trajectories for trapped electrons that receive wave parallel momentum. The dashed black line is a flux surface. For both signs of $v_\|$ at the outer-midplane on the flux surface, the orbits move inward and increase its width with every bounce.
}
\end{figure}

\section{Conclusion}\label{Conclusion}
\noindent In this paper, we have proven that wave-particle momentum transfer by resonance happens through both the parallel motion and the gyro-motion. Only considering the parallel motion leads to incorrect results when evaluating toroidal angular momentum transfer. The toroidal momentum carried by the parallel motion is rapidly dissipated by collisions with ions. The perpendicular force is balanced by an electron radial pinch rather than collisions. For the LH wave in tokamaks, the difference between the toroidal wave vector and the toroidal component of the parallel wave vector, represented by $n_\phi R - n_\|( B_{\phi}/B)R$, determines the radial pinch of the electrons. Typically, the high poloidal wave number at the electron Landau resonance induces a pinch with the opposite sign to the pinch that one would have expected for a wave with no poloidal wave number and the same toroidal wave number. For counter-current direction LH wave momentum input, this pinch is O(1mm/sec) and it generates an additional outward radial electric field that makes the flux of electrons and ions ambipolar. This radial electric field gives an $E \times B$ ion flow with the opposite direction to the momentum source. Eventually, after receiving the parallel momentum by collisions with electrons, the ion velocity will acquire the direction of the momentum source. The ions achieve the toroidal angular momentum that the LH wave contained originally through two main channels: ion-electron collision for the parallel direction motion, and the Lorentz force due to an outward radial ion pinch following the electron pinch.

\section*{Acknowledgments}
We would like to thank Dr. Abhay Ram for useful discussions and insights about lower hybrid waves. 
This work was supported in part by Samsung scholarship and by U.S. Department of Energy Grant No. DE-SC008435, No. DE-FC02-01ER54648, and No. DE-FG02-91ER54109.

\appendix
\section{Wave power absorption by quasilinear diffusion}\label{appendixA}
To evaluate the wave power absorption in (\ref{Q_avg1}), we utilize the Bessel function expansion for the sinusoid phase,
 \begin{eqnarray}
e^{i\lambda \sin {\eta}}&=&\sum_n e^{in\eta} J_n(\lambda),\\
\sin{\eta} e^{i\lambda \sin {\eta}}&=&-\sum_n i e^{in\eta} J_n^{\prime}(\lambda),\\
\cos{\eta} e^{i\lambda \sin {\eta}}&=&\sum_n \frac{n}{\lambda} e^{in\eta} J_n(\lambda),
\end{eqnarray}
and the sifting property of the phase average 
 \begin{eqnarray}
 \fl \;\;\;\;  \int_0^{2\pi} d\eta e^{-i\lambda(  \sin{(\eta+\Omega \tau)}- \sin{(\eta)})} &=&\int_0^{2\pi} d\eta\sum_l e^{-il(\eta+\Omega\tau)} J_l(\lambda) \sum_n e^{in\eta} J_n(\lambda)\\
  \fl \;\;\;\; &=& 2\pi \sum_n e^{-in\Omega\tau} J^2_n(\lambda). \label{sift2}
 \end{eqnarray}
Then, the power absorption $P_{abs,\mathbf{k}}$ for a single mode can be evaluated \cite{Kennel:POF1966,Stix:AIP1992} as
\begin{eqnarray}
  P_{abs}&\equiv&\sum_\mathbf{k}P_{abs,\mathbf{k}}=\int_{-\infty} ^{\infty} dv_{\|} \int_0^{\infty} dv_{\perp} 2\pi v_{\perp}  \left \langle \frac{mv^2}{2} Q(f) \right \rangle_{\alpha} ,\nonumber\\
\end{eqnarray}
giving
\begin{eqnarray}
 \fl P_{abs,\mathbf{k}}&=&\frac{{\pi} Z^2 e^2 }{m^2}   \int_{-\infty} ^{\infty} dv_{\|} \int_0^{\infty} dv_{\perp} 2\pi v_{\perp}   \sum_n \frac{mv^2}{2} L \bigg (v^2_{\perp} \delta (\omega-k_{\|}v_{\|}-n\Omega) |\chi_{\mathbf{k},n}|^2 L(f_0) \bigg ) \label{P_abs1}
\\
 \fl &=&-\frac{{\pi} Z^2 e^2}{m^2} \int_{-\infty} ^{\infty} dv_{\|} \int_0^{\infty} dv_{\perp} 2\pi v_{\perp}  \sum_n m v_{\perp}^2 \delta (\omega-k_{\|}v_{\|}-n\Omega) |\chi_{\mathbf{k},n}|^2 L(f_0).\label{P_abs2}
\end{eqnarray}

\section{Wave momentum transfer by resonances}\label{appendixB}
The toroidal momentum transfer in the parallel direction (\ref{mom_par1}) can be written by inserting the flux $\Gamma_{\|}$ (\ref{Gamma_par}) and the perturbed fluctuated distribution function $f_\mathbf{k}$ (\ref{f_k2}).
\begin{eqnarray}
 \fl P_{\phi}^{\|} &=&ZeR \sum_\mathbf{k}\int_{-\infty} ^{\infty} dv_{\|} \int_0^{\infty} dv_{\perp} v_{\perp} \int_0^{2\pi} d\alpha  ( \mathbf{\hat{z}} \cdot \boldsymbol{\hat{\phi}} ) \nonumber \\\fl &\times&  \bigg\{E^{*}_{\mathbf{k},\|} \bigg(1-\frac{k_{\perp}v_{\perp}}{\omega} \cos{(\alpha-\beta)} \bigg) + E^{*}_{\mathbf{k},\perp}\frac{k_{\|}v_{\perp}}{\omega}\bigg \}{f}_\mathbf{k}\nonumber\\
 \fl &=&-\frac{Z^2e^2 R}{m} \sum_\mathbf{k}\int_{-\infty} ^{\infty} dv_{\|} \int_0^{\infty} dv_{\perp} 2\pi v_{\perp} \int_0^{\infty} d\tau  e^{i(\omega-k_{\|}v_{\|})\tau}  \int_0^{2\pi} d\eta( \mathbf{\hat{z}} \cdot \boldsymbol{\hat{\phi}} ) \nonumber\\ 
 \fl &\times&  \sum_n e^{in\eta} \left(\left(1-\frac{n\Omega}{\omega}\right)J_n E^{*}_{\mathbf{k},\|} + \frac{nJ_n}{\lambda}\frac{k_{\|}v_{\perp}}{\omega}( E^{*}_{\mathbf{k},+}+E^{*}_{\mathbf{k},-})+J_n^{\prime}\frac{k_{\|}v_{\perp}}{\omega}( E^{*}_{\mathbf{k},+}-E^{*}_{\mathbf{k},-})  \right)   \label{P_par1}\nonumber\\
 \fl &\times&\sum_l e^{-il(\eta+\Omega\tau)} \left(E_{\mathbf{k},\|} \frac{\partial f_0}{\partial v_{\|}}J_l+\frac{lJ_l}{\lambda}(( E_{\mathbf{k},+}+E_{\mathbf{k},-})U-E_{\|}V)+( E_{\mathbf{k},+}-E_{\mathbf{k},-})U J^{\prime}_l   \right) \nonumber. 
\end{eqnarray}
The phase $i(\omega-k_{\|}v_{\|}-n\Omega)\tau$ is averaged out in the $\tau$ integration except where $\omega-k_{\|}v_{\|}-n\Omega=0$. Using the Dirac-delta function to express this resonance condition, the momentum transfer becomes
\begin{eqnarray}
 \fl P_{\phi}^{\|}&=  -\frac{{\pi} Z^2e^2 R}{m} \sum_\mathbf{k} \int_{-\infty} ^{\infty} dv_{\|} \int_0^{\infty} dv_{\perp} 2\pi v_{\perp}   \sum_n \delta (\omega-k_{\|}v_{\|}-n\Omega) ( \mathbf{\hat{z}} \cdot \boldsymbol{\hat{\phi}} )\nonumber\\ 
 \fl &\times \left(\left(1-\frac{n\Omega}{\omega}\right)J_n E^{*}_{\mathbf{k},\|} + \frac{nJ_n}{\lambda}\frac{k_{\|}v_{\perp}}{\omega}( E^{*}_{\mathbf{k},+}+E^{*}_{\mathbf{k},-})+J_n^{\prime}\frac{k_{\|}v_{\perp}}{\omega}( E^{*}_{\mathbf{k},+}-E^{*}_{\mathbf{k},-})  \right)   
\nonumber\\ 
 \fl &\times \left(E_{\mathbf{k},\|} \frac{\partial f_0}{\partial v_{\|}}J_n+ \frac{nJ_n}{\lambda}(( E_{\mathbf{k},+}+E_{\mathbf{k},-})U-E_{\mathbf{k},\|}V)+( E_{\mathbf{k},+}-E_{\mathbf{k},-})U J^{\prime}_n   \right)\label{P_par1}\\ 
 \fl =& - \frac{{\pi} Z^2e^2 R}{m}\sum_\mathbf{k}  \int_{-\infty} ^{\infty} dv_{\|} \int_0^{\infty} dv_{\perp} 2\pi v_{\perp}    \nonumber \\\fl &\times \sum_n \delta (\omega-k_{\|}v_{\|}-n\Omega) ( \mathbf{\hat{z}} \cdot \boldsymbol{\hat{\phi}} ) \frac{k_{\|}v^2_{\perp}}{\omega} |\chi_{\mathbf{k},n}|^2 L(f_0)\label{P_par2}
\\
 \fl   =& \sum_\mathbf{k}\frac{k_\|}{\omega}P_{abs,\mathbf{k}} R( \mathbf{\hat{z}} \cdot \boldsymbol{\hat{\phi}} )=  \sum_\mathbf{k}\frac{n_\|}{c}P_{abs,\mathbf{k}} R( \mathbf{\hat{z}} \cdot \boldsymbol{\hat{\phi}} )\label{P_par5},
\end{eqnarray}
where  the resonance condition $\omega-k_{\|}v_{\|}-n\Omega=0$, the Bessel function identities $n J_n/\lambda= (J_{n+1}+J_{n-1})/2$ and $J_n^\prime= (J_{n-1}-J_{n+1})/2$, and $\chi_{\mathbf{k} ,n}=E_{\mathbf{k} ,\|} J_{n}\frac{v_{\|}}{v_{\perp}}+  E_{\mathbf{k} ,+} J_{n-1} +E_{\mathbf{k} ,-} J_{n+1}$ are used from (\ref{P_par1}) to (\ref{P_par2}).\\ \\
The rest of the toroidal momentum transfer in (\ref{mom_perp0}) can be obtained by inserting the perturbed fluctuated distribution function $f_\mathbf{k}$ (\ref{f_k2}). Before doing so, we rewrite (\ref{mom_perp0}) as
 \begin{eqnarray}
 \fl&\Delta P_{\phi}^{\perp}+\Delta P_{\phi}^{\alpha} =ZeR\sum_\mathbf{k}\int_{-\infty} ^{\infty} dv_{\|} \int_0^{\infty} dv_{\perp} 2\pi v_{\perp} \int_0^{2\pi} \frac{d\alpha}{2\pi}\nonumber\\ 
 \fl &\times
  \bigg [E^{*}_{\mathbf{k},\|} \frac{k_{\perp}v_{\|}}{\omega} (\cos{\beta} ( \mathbf{\hat{x}} \cdot \boldsymbol{\hat{\phi}} )+ \sin{\beta} ( \mathbf{\hat{y}} \cdot \boldsymbol{\hat{\phi}} ))\nonumber\\
 \fl   &+\bigg(1-\frac{k_{\|}v_{\|}}{\omega} \bigg)\bigg\{\bigg((E^{*}_{\mathbf{k},+}+E^{*}_{\mathbf{k},-})\cos{\beta}-i(E^{*}_{\mathbf{k},+}-E^{*}_{\mathbf{k},-})\sin{\beta}\bigg)(\mathbf{\hat{x}} \cdot \boldsymbol{\hat{\phi}}) \nonumber\\
 \fl   &+\bigg((E^{*}_{\mathbf{k},+}+E^{*}_{\mathbf{k},-})\sin{\beta}+i(E^{*}_{\mathbf{k},+}-E^{*}_{\mathbf{k},-})\cos{\beta} \bigg)(\mathbf{\hat{y}} \cdot \boldsymbol{\hat{\phi}})  \bigg\}
+i \bigg( \frac{k_{\perp}v_{\perp}}{\omega}\bigg)(E^{*}_{\mathbf{k},+}-E^{*}_{\mathbf{k},-}) \nonumber\\ \fl   &\times\bigg\{(\sin{\eta}\cos{\beta}+\cos{\eta}\sin{\beta})  ( \mathbf{\hat{x}} \cdot \boldsymbol{\hat{\phi}} )-( \cos{\eta}\cos{\beta}-\sin{\eta}\sin{\beta})( \mathbf{\hat{y}} \cdot \boldsymbol{\hat{\phi}} ) \bigg\}  \bigg] f_\mathbf{k} \label{P_add2}. 
  \end{eqnarray}
 Using (\ref{f_k2}) for $f_\mathbf{k}$ and the Dirac-delta function for the resonance condition gives
\begin{eqnarray}
 \fl&\Delta P_{\phi}^{\perp}+\Delta P_{\phi}^{\alpha} \nonumber \\ \fl =&  -\frac{{\pi} Z^2e^2 R}{m} \sum_\mathbf{k} \int_{-\infty} ^{\infty} dv_{\|} \int_0^{\infty} dv_{\perp} 2\pi v_{\perp}  \delta (\omega-k_{\|}v_{\|}-n\Omega)  \bigg[ (\cos{\beta} ( \mathbf{\hat{x}} \cdot \boldsymbol{\hat{\phi}} )+ \sin{\beta} ( \mathbf{\hat{y}} \cdot \boldsymbol{\hat{\phi}} ))\nonumber\\ 
 \fl&\times \left\{\frac{k_{\perp}v_{\|}}{\omega}J_n E^{*}_{\mathbf{k},\|} + \bigg(1-\frac{k_{\|}v_{\|}}{\omega} \bigg)J_n(E^{*}_{\mathbf{k},+}+E^{*}_{\mathbf{k},-})+J_n^{\prime}\frac{k_{\perp}v_{\perp}}{\omega}(E^{*}_{\mathbf{k},+}-E^{*}_{\mathbf{k},-})  \right\}\nonumber\\
 \fl&-i(\sin{\beta} ( \mathbf{\hat{x}} \cdot \boldsymbol{\hat{\phi}} )- \cos{\beta} ( \mathbf{\hat{y}} \cdot \boldsymbol{\hat{\phi}} )) )\nonumber\\ 
 \fl &\times \left\{\bigg(1-\frac{k_{\|}v_{\|}}{\omega} -\frac{n\Omega}{\omega} \bigg)J_n(E^{*}_{\mathbf{k},+}-E^{*}_{\mathbf{k},-}) \right\}
\bigg]\chi_{\mathbf{k},n} v_{\perp}L(f_0) \label{P_add3_a}\\
 \fl=&-  \frac{{\pi} Z^2e^2 R}{m}  \sum_\mathbf{k}\int_{-\infty} ^{\infty} dv_{\|} \int_0^{\infty} dv_{\perp} 2\pi v_{\perp}   \sum_n \delta (\omega-k_{\|}v_{\|}-n\Omega)\nonumber\\ 
 \fl&\times  \left(\cos{\beta} ( \mathbf{\hat{x}} \cdot \boldsymbol{\hat{\phi}} )+ \sin{\beta} ( \mathbf{\hat{y}} \cdot \boldsymbol{\hat{\phi}} )\right) \frac{k_{\perp}v^2_{\perp}}{\omega} |\chi_{\mathbf{k},n}|^2 L(f_0)\label{P_add5_a}
\\
  \fl=&  \sum_\mathbf{k}\frac{\mathbf{k_{\perp}} \cdot {\boldsymbol{\hat{\phi}}}}{\omega}P_{abs,\mathbf{k}} R=  \sum_\mathbf{k}\frac{n_\perp}{c}P_{abs,\mathbf{k}} R  \left(\cos{\beta} ( \mathbf{\hat{x}} \cdot \boldsymbol{\hat{\phi}} )+ \sin{\beta} ( \mathbf{\hat{y}} \cdot \boldsymbol{\hat{\phi}} )\right)
\label{P_add6}.
\end{eqnarray}
From step (\ref{P_add3_a}) to (\ref{P_add5_a}), the resonance condition and the Bessel function identities $n J_n/\lambda= (J_{n+1}+J_{n-1})/2$ and $J_n^\prime= (J_{n-1}-J_{n+1})/2$ are used.

\section{Bounce averaging and flux surface averaging}\label{appendixC}
\noindent The bounce-average is defined as $\langle X \rangle_{b}  = \frac{1}{\tau_b} \oint \frac{d\ell \; X}{v_{\|}}=\frac{1}{\tau_b} \oint \frac{d\theta \; X}{v_{\|}\mathbf{\hat{z}} \cdot \nabla\theta}$,  where $\theta$ is the poloidal angle and $\tau_b  (\mathcal{E},\mu) =  \oint   \frac{d\ell} {v_{\|} (\mathcal{E},\mu, l)} $ is the bounce time. The flux surface average is $\langle X \rangle_{s}= \frac{1}{\tau_s} \oint \frac{d\theta \; X}{\mathbf{B} \cdot \nabla\theta}=\frac{1}{\tau_s} \oint \frac{d\ell \; X}{B} $, because the Jacobian is $J=\frac{1}{\nabla\phi \times \nabla\psi \cdot \nabla\theta}=\frac{1}{\mathbf{B} \cdot \nabla\theta}$. The normalization factor is $\tau_s  =  \oint   \frac{d\ell}{B} $. There is a relation between the flux surface averaged value and the bounce averaged value for passing particles,
\begin{eqnarray}
\langle X \rangle_{b}\tau_b  = \left \langle \frac{B}{v_{\|}}X \right \rangle_{s} \tau_s.\label{b_vs_s1}
\end{eqnarray}
For trapped particles, assuming that the flux surface average is taken between the turning points of the orbit, the bounce averaged value is defined as
\begin{eqnarray}
\langle X \rangle_{b} \equiv \frac{1}{2}\sum_\sigma \frac{\tau_{s\frac{1}{2}} }{\tau_{b\frac{1}{2}} } \left \langle \frac{B}{\vert v_{\|}\vert}X \right \rangle_{s\frac{1}{2}}, \label{b_vs_s2}
\end{eqnarray}
where $\tau_{s\frac{1}{2}}$ and $\tau_{b\frac{1}{2}}$ stand for the integration and the bounce time from one turning point to the other turning point, and the summation over $\sigma=v_\|/|v_\||$ indicates that the values of $X$ for the two parallel velocity signs must be added. Bounce averaging annihilates the operator $v_{\|}\nabla_{\|}X$ for both passing and trapped particles.\\

\end{document}